\documentclass[sigconf, authorversion=true]{acmart}
\pdfoutput=1
\usepackage[english]{babel}
\usepackage{booktabs} % For formal tables
\usepackage{multirow}

\usepackage{parskip}
\usepackage{balance}

\usepackage{tikz}
\usetikzlibrary{matrix}

\newtheoremstyle{techniques}%
{0.5em}% space above
{0em}% space below
{\itshape}% body font
{1em}% indent amount
{}% head font
{}% punctuation after head
{1em}% space after head
{\thmname{#1}\thmnumber{#2}.}% head spec
\theoremstyle{techniques}

\newtheorem{technique}{$\tau$}

\copyrightyear{2018}
\acmYear{2018}
\setcopyright{acmcopyright}
\acmConference[SAICSIT '18]{2018 Annual Conference of the South African Institute of Computer Scientists and Information Technologists}{September 26--28, 2018}{Port Elizabeth, South Africa}
\acmBooktitle{2018 Annual Conference of the South African Institute of Computer Scientists and Information Technologists (SAICSIT '18), September 26--28, 2018, Port Elizabeth, South Africa}
\acmPrice{15.00}
\acmDOI{10.1145/3278681.3278703}
\acmISBN{978-1-4503-6647-2/18/09}

\begin{document}

\title{Deploying South African Social Honeypots on Twitter}
%\titlenote{}
%\subtitle{Extended Abstract}
%\subtitlenote{The full version of the author's guide is available as
  %\texttt{acmart.pdf} document}

\author{Laurenz A. Cornelissen}
%\authornote{}
\orcid{0000-0001-7864-3143}
\affiliation{%
  \institution{Computational Social Science Group\\
  Centre for AI Research\\
  Department of Information Science\\
  Stellenbosch University}
  }
\email{alducornelissen@sun.ac.za}

\author{Richard J Barnett}
\orcid{0000-0003-4133-4587}
\affiliation{%
  \institution{Computational Social Science Group\\
  Centre for AI Research\\
  Department of Information Science\\
  Stellenbosch University}
  }
\email{barnettrj@acm.org}

\author{Morakane A.M. Kepa}
\orcid{0000-0002-3923-5224}
\affiliation{%
  \institution{Computational Social Science Group\\
  Department of Information Science\\
  Stellenbosch University}
  }
\email{morakanekepak@gmail.com}

\author{Daniel Loebenberg-Novitzkas}
\orcid{0000-0003-1947-9990}
\affiliation{%
  \institution{Computational Social Science Group\\
  Department of Information Science\\
  Stellenbosch University}
  }
%\email{19367309@sun.ac.za}

\author{Jacques Jordaan}
\orcid{0000-0001-7864-3143}
\affiliation{%
  \institution{Computational Social Science Group\\
  Department of Information Science\\
  Stellenbosch University}
  }
%\email{15447928@sun.ac.za}

% % The default list of authors is too long for headers.
 \renewcommand{\shortauthors}{L.A. Cornelissen \emph{et al.}}

\begin{abstract}
Inspired by the simple, yet effective, method of tweeting gibberish to attract automated social agents (bots), we attempt to create localised honeypots in the South African political context. We produce a series of defined techniques and combine them to generate interactions from users on Twitter. The paper offers two key contributions. Conceptually, an argument is made that honeypots should not be confused for bot detection methods, but are rather methods to capture low-quality users. Secondly, we successfully generate a list of 288 local low quality users active in the political context.
\end{abstract}




\maketitle\newpage

\section{Introduction}
\label{S:1}
The prevalence of automated social agents on online social networks (OSNs) increasingly necessitate research on the phenomenon. An automated agent, colloquially referred to as a bot on OSN platforms such as Twitter, is a program which performs scripted actions such as following, tweeting, favouriting or retweeting. These automated agents vary considerably in their uses and implementations. Such automated agents have high utility for organisations and even individual users. Organisations increasingly find use for automated agents in client relations management to improve their capacity to deal with a large number of repetitive queries by clients. Many examples of automated agents exist, which are aimed at performing scripted utility functions for individuals. A simple example is a social media assistant, which reacts to queries of users or to unfold a thread on Twitter. Individuals can also augment their profiles by using automated scripts, which turns them into what is usually labelled a cyborg. Cyborgs therefore exhibit many features of automated agents, but are linked to a specific human user.

The use of automated social agents on OSNs is, therefore, not a new phenomenon, but there is a recent increase in their application for political purposes \citep{Howard2018}. These automated agents come in different guises. The majority of bots and use-cases are trivial to implement and have a negligible effect on the average user. There has, however, been a rise in the use of automated agents to effect large scale influence according to political agendas, specifically on social media. These political automated agents are used in a number of ways. \citet{Howard2018} explores some of the specifics of political bot activities, especially as observed during the 2016 United States presidential election. This particular political event is widely referenced as involving a notable increase in such activity. The genesis, of this type, and level, of interference in Western democratic politics might be the British referendum on their European Union membership, colloquially known as Brexit. \citet{Howard2016} discovered that up to 32\% of the conversation on Twitter, involving the referendum, may have been driven by political bots. \citet{Bastos2017} investigated the role of these political bots in sharing partisan and inaccurate news. In a review of the data of the referendum, \citet{Narayanan2017} suggests evidence of involvement from Russian borne actors in the creation of political bots. The same patterns have been observed during the 2016 US presidential election, especially an increase in the use of Facebook as a platform for political interference \citep{Howard2018}.

South Africa has not been spared in this new practice of political interference. There are observed cases of political bots on Twitter, and an increase in partisan or false news websites. These revelations have mostly been as the result of individual journalists investigations into suspicions activity involving a particular organisation's role in this practice \citep{VanNiekerk2018}. South Africa is facing its first national elections since this new phenomenon. As the economic centre of Sub-Saharan Africa, there is enough political interest to attract such political interference campaigns. There is, therefore, a need to develop a means of systematically investigating political bots within the South African political context.

Below, we offer a brief literature review to highlight prior advances in this field and some key problems. Based on these problems, we propose a systematic approach to developing a honeypot within a specific context, which should be able to attract political bots which are created for political engagement in South Africa.

\section{Literature Review}
To develop effective social honeypots on Twitter, we build on prior research. We only look at literature which deployed a set of honeypots on Twitter. We are interested in their honeypot design procedures, specifically which activities are more successful than others in attracting active automated agents and other low quality users within a particular context.

\subsection{Honeypots On Twitter}

The term honeypot has its origins in network security research, which defines it as a decoy computer resource used in networks to deter system attackers from the real system \citep{Mokube2007}. The main aim of designing a honeypot is to have the honeypot compromised, probed or attacked. Through monitoring the honeypot, researchers become more aware of activities that attackers perform when attacking systems. This allows for better network design and security measures.

The same method is applied to OSNs, where a user profile is created with the purpose of attracting interaction from attackers on the network. By recording the interactions with the honeypot, itself also an automated agent, the researcher compiles a record of profiles which are potentially automated agents. This was the approach followed by \citet{Lee2011}; they created automated agents which tweeted gibberish, with the aim of attracting other automated agent\-s, while avoiding humans. More recently, \citet{Morstatter2016} defined social honeypots as automated accounts that are used to lure other automated accounts by exhibiting non-human behaviours.

\citet{Lee2011} deployed 60 social honeypots on Twitter for 7 months. The social honeypots were divided into 4 types based on how they tweet. The first tweeted normal text, the second used replies to other social honeypots. The third type tweeted only URLs, while the fourth tweeted containing trending Twitter topics. All of these honeypots tweeted in gibberish, with the intention of only attracting automated agents, which are not capable of distinguishing between gibberish and non-gibberish content. They recorded 36000 interactions over the seven months.

There are multiple research efforts extending this initial attempt at designing social honeypots on OSNs. The next section offers a brief overview of the designs of prior research. We then critique these design methods and provide a resolution before presenting our own methodology for developing social honeypots within a specific context.

\subsection{Honeypot Designs}
This section specifically reviews previous honeypot design efforts, in order to identify successful designs to implement within this new context.

\citet{Lee2010} deployed honeypots on two OSNs, Myspace and Twitter. Since we are interested in Twitter, we only report their honeypot design on Twitter. They deployed a mixed design of honeypots containing two features: whether they had profile information and whether they had tweeted. They deployed the honeypots for a month, during which they received interaction from 500 profiles. The same dataset was used for \citet{Lee2010a}, which offers no extra explanation as to their social honeypot design.

\citet{Lee2011} offered more specifics regarding their social honeypot methodology. They created 60 social honeypots, which were deployed on Twitter for seven months. Each honeypot could post four different types of tweets; a normal textual tweet which the researchers do not define; an `@' reply to other honeypots; a tweet containing a URL link; or, a tweet containing one of Twitter's current top ten trending topics. By having four different types of tweets it brings attention to the importance of tweet type when attracting automated agents.

\citet{Yang2014} offered a more detailed design. They created 96 social honeypots, which they refer to as benchmarks. They also identified three types of behaviours on Twitter as: tweeting, following, and application platform. For each behaviour, distinct patterns were identified.

Under tweet behaviour, the three patterns were: tweet frequency, tweet keywords, and tweet topics. For tweet frequency the researchers found that social honeypots that tweeted more frequently attracted more automated agents. On tweet keywords the choices were on using: popular trending topics; random hashtags; tweets on current affairs; bait words; and tweets containing no hashtags. From these actions, tweeting on trending topics and tweets with hashtags attracted the most accounts, while current affairs attracted the least. This highlights the sensitivity of interaction based on tweet content type. Lastly, to understand the behaviour of following actions, five social honeypots followed two verified accounts, of people in various fields, a day. From these, those following users in entertainment attracted the most automated agents.

\citet{Elmendili2017} deployed twenty social honeypots on Twitter, which attracted interactions from 300 accounts, of which 90 were confirmed to be automated agents. Unfortunately, they do not elaborate on the design of the honeypots.

\citet{Morstatter2016} offered more information on the design of their social honeypots. They developed nine honeypots which tw\-eeted Arabic phrases. Each honeypot randomly followed other honeypots, randomly retweeted other honeypots, and retweeted the owners of the Arabic phrases. The use of a language feature is a unique contribution to the design of social honeypots. Their social honeypots attracted interactions from 3602 accounts, which they assume to be bots. This assumption is common, however, as \citet{Wang2012} highlighted; automated agents are able to comprehend statues on Twitter. This weakens the assumption that bots would not be able to discriminate between other bots, tweeting gibberish, and humans. Moreover, \citet{Clark2016} introduces another class of user: \textit{cyborgs}. Cyborgs are  either humans who use algorithms to alter their profiles, or are automated agents that are run directly by a human. This effectively weakens the assumption further, because the honeypot might be attracting these cyborgs who are actually humans. This supports the argument that there is a range of poor quality users on social media platforms. Poor quality, in this instance, relates to the lack of social authenticity of a user. For instance, a cyborg that automatically follows other users back is not an honest social reaction to being followed.

From the literature above, it is clear that, apart from \citet{Yang2014} there is a lack of systematic design elements to the honeypots themselves. The activities of the honeypots are unclear or uninformative for external researchers. \citet{Lee2010} gave incomplete information about the profiles of their honeypots. \citet{Lee2011} give no explanation of how frequently their honeypots tweet. \citet{Elmendili2017} gives no explanation about the profiles of their honeypots just that they had legitimate profiles which they do not describe. Like \citet{Lee2010} and \citet{Lee2011} they also do not explain how frequently their honeypots tweeted.

Based on the review of the literature it is unclear which methods are more effective than others in either attracting interactions from other profiles, or how effective the claim is of only attracting bots. When the effectiveness of attracting only automated agents was actually measured, it turns out to be 30\% of the attracted profiles. The method is therefore not an effective bot classification methodology. Although classification should not be the primary objective of social honeypots, the narrative within the literature either assumes this, like \citet{Morstatter2016}, or it is used only as a step in classification methodologies.

We, therefore, have two objectives. Firstly, we argue that the objective of a social honeypot is not to classify automated agents, but rather detect candidates for classification. Secondly, we offer a detailed methodology of social honeypot behaviour on Twitter by designing specific techniques. Paired with our overall objective of discovering automated agents within a particular political and geographical context, we then present our results.

The next section outlines our proposed distinction between classification and detection objectives in automated agent research.

\section{Classification versus Detection}
We propose a clarification of the distinction between classification and detection objectives in automated agent research. The distinction is important because it determines the success criteria for the methodologies.

Classification methodologies are concerned with the classification of a user as belonging to a particular class. In most cases, these classes are either automated agents or humans. Most bot detection research methods are concerned with classification.

To develop classification methodologies researchers require an annotated dataset of users which provides the ground truth to train models on the features of the classes. These trained models are then capable of classifying unlabelled data into the classes by analy\-sing multiple features of each user. Whether the objective is to classify, or to develop a classifier, the procedure requires data, it does not necessarily discover data.

In contrast, detection methodologies should focus on the discovery of data. There are various detection methods, of which social honeypots is only one example. For instance, to gather user profiles of legitimate users, instead of automated agents, \citet{Morstatter2016} manually selected ten legitimate users who tweeted particular Arabic phrases to generate a seed list of profiles. The friends of these profiles were collected using snowball sampling. Social honeypots are, therefore, more focussed on attracting users which are of interest to be classified, and should therefore be labelled a detection method, since it detects possible users of interest, and not necessarily classifies them.

A helpful framework to introduce is the distinction between precision and recall, where classification is concerned with precision, and detection with recall.

Precision is a measure of how accurately a method can discriminate between classes, while minimising false positives. When a classifier is used on OSNs, false positives are humans being detected as bots, if the classification leads to a suspension of the user the OSN would suffer disgruntled users. Thus they would avoid false positives. In contrast, recall aims to reduce false negatives, i.e. if there is an automated agent, the method should record it, regardless of falsely recording humans in the process.

In classification it is common to qualify the performance of the classifiers according to their precision, or a number of other related performance measures based off a confusion table. It is, however, not possible to accurately measure the recall ability of a social honeypot, since the number of existing automated agents are always unknown. The only way to accurately measure it is in controlled environments where the number of target agents are known. \citet{Morstatter2016} also found that most social honeypot research favours precision above recall.

A honeypot methodology should, therefore, prioritise recall, which is to attract as many automated agents as possible. It is not the intention of a honeypot to develop robust classifiers or filter out unknown spammers. To improve recall, the methodology should be able to distinguish between honeypot activities, which are more effective at generating interactions within certain contexts, and should be able to detect a wide assortment of agents.

Previous literature has shown that from the attracted agents, only about 30\% can be classified as actual automated agents. The other 70\% of attracted agents are not, however, false positives. They should rather be regarded as poor quality agents for further investigation. Currently, classifiers are only interested in automated agents that lead to a bias, which ignores other poor quality users. There is some nuance introduced in different type of automated agents, such as political bots or spamming bots \citep{Varol2017}. Honeypots should, therefore, be aimed at attracting poor quality agents of which a subset may be automated agents.

\section{Methodology}

In order to attract active automated agents which are active in the South African political sphere, we have developed a series of honeypots based on combinations of identified techniques. The overall objective is to identify a combination of simple techniques which maximise the number of automated agent followers who engage with the honeypots. In order to achieve this, we have employed the following methodology.

Initially we used literature to identify a number of simple techniques which may be expected to attract automated agents, these techniques were then tested, individually and in combinations again\-st Twitter for a period of time before being evaluated for their effectiveness. This section discusses the approach taken in the identification of techniques and testing the effectiveness of each technique.

\subsection{Solicitation Techniques}

The techniques for attracting automated agents can be classified into two very broad areas, those generating their own content, and those relying on interactions with content generated by other users. We define these techniques in this section with the $\tau$ symbol.

\subsubsection{Content Interaction}

Twitter permits three general interactions with users and statuses (without generating new content). These are \emph{friending} a user, as well as \emph{favouriting} or \emph{retweeting} a status. The techniques used to rely on other users are, therefore, limited to these three broad categories. In order to solicit automated agents based on content interactions, we must, therefore, identify users to \emph{friend} and statuses to \emph{favourite} or \emph{retweet}. Since this research focuses on a South African political context, we focus on users who are political actors in South Africa.

Political actors were identified in a two step process. First the websites of the thirteen political parties were consulted to obtain a list of party leaders. Some parties linked to Twitter profiles for leaders, while we searched Twitter for others. The list of political leaders was augmented by searching news articles for other relevant persons and then evaluating their Twitter profiles to confirm relevance. This process resulted in a total of 287 political actors, inclusive of 44 political leaders. Direct interaction with these political actors results in the first three techniques for attracting automated agents.

\begin{technique}Follow a series of political actors.\end{technique}
\begin{technique}Favourite the most recent status of a series of political actors.\end{technique}
\begin{technique}Retweet the most recent status of a series of political actors.\end{technique}
%\begin{technique}\end{technique}

In addition to the most recent status of these users, we also identified techniques which rely on content generated by these users. This effectively identifies additional statuses and users to interact with. A list of \emph{trending} keywords and key phrases was generated from the 44 political leaders, based on their activity in the previous week.\footnote{These keywords and key phrases were generated independently of the Twitter feature of the same name. This is because, we only wished to identify trends amongst these specific political actors; and secondly, because the use of Twitter trends by automated agents is against the Twitter terms of service.} These keywords and key phrases were ranked based on their popularity amongst the users. We then used these keywords to identify six additional techniques based on interacting with the list sequentially or in a random fashion. As with the keywords and key phrases themselves, the content investigated by these techniques was generated over the proceeding week. This allowed the content to be relevant but not real-time.

\begin{technique}Follow users that post statuses about a keyword topic in a sequential way.\end{technique}
\begin{technique}Follow users that post statuses about a keyword topic in a random way.\end{technique}
\begin{technique}Favourite a status about a keyword topic in a sequential way.\end{technique}
\begin{technique}Favourite a status about a keyword topic in a random way.\end{technique}
\begin{technique}Retweet a status about a keyword topic in a sequential way.\end{technique}
\begin{technique}Retweet a status about a keyword topic in a random way.\end{technique}

Finally for techniques that consume content, Twitter permits users to geotag their statuses with either a location or a place. Twitter also permits real-time access to a subset of all statuses produced on the platform or filtered based on a series of criteria. We used these features to look at real-time statuses on the topics identified by the keywords and key phrases above and were also geotagged as being from or within South Africa. These statuses are then interacted with in the same three ways as with the other techniques.

\begin{technique}Follow users that post geo-local statuses about a real-time keyword topic.\end{technique}
\begin{technique}Favourite geo-local statuses about a real-time keyword topic.\end{technique}
\begin{technique}Retweet geo-local statuses about a real-time keyword topic.\end{technique}

\subsubsection{Content Generation}

The second approach to attracting automated agents is to generate new content as statuses on the platform. For the purposes of this research we do not define \textit{new} as original content, but rather content, which forms its own independent status.

In order to post statuses that are likely to attract automated agents, we have identified two areas to obtain relevant content from. The first of these areas is content that has already been generated by the list of 287 political actors previously identified. Whilst this content is essentially duplicated from these users, it is posted as new statuses rather than as interactions with existing statuses. For the purposes of the research, the content of these 287 users from the proceeding seven days was used. The second source of politically relevant content is news. To this end we selected the headlines and abstracts of news articles from the RSS feeds of both News24 and Google News. Only South African news stories are used in both cases. As news can progress somewhat rapidly, only the proceeding day's news is used at any given time.

As South Africa is linguistically diverse, with eleven official languages, we looked to produce content in languages other than English. All the news content provided by both News24 and Google is in English and the vast majority of observed statuses on Twitter are also in English. We used Google Translate to convert all the statuses and news articles into Afrikaans, Xhosa and Zulu.

Some research shows that automated agents do not look at sentence structure or at full sentences when consuming content. It is assumed that human agents will not friend a user who posts statuses that are not coherent sentences. For this reason, we also randomise the word order of all sources, across all languages.

Finally, in selecting content to produce, we take cognisance that Twitter is an online social network and as a result, we include mentions to one of the 287 political actors as well as hashtags to our locally identified trending keywords.

% \usepackage{booktabs}
% \usepackage{multirow}
\begin{table*}[]
	\centering
	\caption{Content Generating Techniques}
	\label{table:cgt}
	\resizebox{\textwidth}{!}{% % Richard, screw you, this is how it is.
		\begin{tabular}{@{}lllllllllllllllllllllllllll@{}}\toprule & & \rotatebox{90}{\stepcounter{technique}$\tau$\arabic{technique}.} & \rotatebox{90}{\stepcounter{technique}$\tau$\arabic{technique}.} & \rotatebox{90}{\stepcounter{technique}$\tau$\arabic{technique}.} & \rotatebox{90}{\stepcounter{technique}$\tau$\arabic{technique}.} & \rotatebox{90}{\stepcounter{technique}$\tau$\arabic{technique}.} & \rotatebox{90}{\stepcounter{technique}$\tau$\arabic{technique}.} & \rotatebox{90}{\stepcounter{technique}$\tau$\arabic{technique}.} & \rotatebox{90}{\stepcounter{technique}$\tau$\arabic{technique}.} & \rotatebox{90}{\stepcounter{technique}$\tau$\arabic{technique}.} & \rotatebox{90}{\stepcounter{technique}$\tau$\arabic{technique}.} & \rotatebox{90}{\stepcounter{technique}$\tau$\arabic{technique}.} & \rotatebox{90}{\stepcounter{technique}$\tau$\arabic{technique}.} & \rotatebox{90}{\stepcounter{technique}$\tau$\arabic{technique}.} & \rotatebox{90}{\stepcounter{technique}$\tau$\arabic{technique}.} & \rotatebox{90}{\stepcounter{technique}$\tau$\arabic{technique}.} & \rotatebox{90}{\stepcounter{technique}$\tau$\arabic{technique}.} & \rotatebox{90}{\stepcounter{technique}$\tau$\arabic{technique}.} & \rotatebox{90}{\stepcounter{technique}$\tau$\arabic{technique}.} & \rotatebox{90}{\stepcounter{technique}$\tau$\arabic{technique}.} & \rotatebox{90}{\stepcounter{technique}$\tau$\arabic{technique}.} & \rotatebox{90}{\stepcounter{technique}$\tau$\arabic{technique}.} & \rotatebox{90}{\stepcounter{technique}$\tau$\arabic{technique}.} & \rotatebox{90}{\stepcounter{technique}$\tau$\arabic{technique}.} & \rotatebox{90}{\stepcounter{technique}$\tau$\arabic{technique}.} & \rotatebox{90}{\stepcounter{technique}$\tau$\arabic{technique}.} \\\midrule
			\multirow{2}{*}{Content Source} & Political Actors & $\times$ & $\times$ & $\times$ & $\times$ & $\times$ & $\times$ & $\times$ & $\times$ & $\times$ & $\times$ & & & & & & & & & & & & & & & \\
			& News & & & & & & & & & & & $\times$ & $\times$ & $\times$ & $\times$ & $\times$ & $\times$ & $\times$ & $\times$ & $\times$ & $\times$ & $\times$ & $\times$ & $\times$ & $\times$ & $\times$ \\\cmidrule{1-2}
			\multirow{4}{*}{Language} & English & $\times$ & $\times$ & $\times$ & & & & & & & & $\times$ & $\times$ & $\times$ & $\times$ & $\times$ & $\times$ & & & & & & & & & \\
			& Afrikaans & & & & $\times$ & $\times$ & & & & & & & & & & & & $\times$ & $\times$ & $\times$ & $\times$ & & & & & \\
			& Xhosa & & & & & & $\times$ & $\times$ & $\times$ & & & & & & & & & & & & & $\times$ & $\times$ & & & \\
			& Zulu & & & & & & & & & $\times$ & $\times$ & & & & & & & & & & & & & $\times$ & $\times$ & $\times$ \\\cmidrule{1-2}
			\multirow{2}{*}{Word Order} & Sentence Order & $\times$ & & & $\times$ & & $\times$ & & & $\times$ & & $\times$ & $\times$ & $\times$ & & & & $\times$ & $\times$ & & & $\times$ & & $\times$ & & \\
			& Randomised & & $\times$ & $\times$ & & $\times$ & & $\times$ & $\times$ & & $\times$ & & & & $\times$ & $\times$ & $\times$ & & & $\times$ & $\times$ & & $\times$ & & $\times$ & $\times$ \\\cmidrule{1-2}
			\multirow{2}{*}{Social Features} & Mentions & & $\times$ & & & & $\times$ & $\times$ & & $\times$ & $\times$ & $\times$ & & & $\times$ & & & $\times$ & & $\times$ & & $\times$ & $\times$ & & & \\
			& Hashtags & & & $\times$ & $\times$ & & & $\times$ & $\times$ & & $\times$ & $\times$ & $\times$ & & & $\times$ & & & $\times$ & $\times$ & & & $\times$ & $\times$ & $\times$ & \\
			\bottomrule
		\end{tabular}%
	}
\end{table*}

Each of the categories in this section are permuted into a variety of different techniques for content generation. Table \ref{table:cgt} shows the permutations we used to generate individual techniques. Not every permutation (a total of 64 permutations exist) was considered. The permutations that were considered were done so due to capacity limitations (total number of Twitter \emph{users} we had access to) and focused on English (as most content on Twitter, even in South Africa is in English) with an approximately even split on the remaining parameters. This results in a total of 25 content generating techniques which we then combined with the consumption techniques.

\subsection{Honeypot Construction}

\begin{figure*}
	\centering
	\begin{tikzpicture}
	\tikzset{square matrix/.style={
			matrix of nodes,
			column sep=-\pgflinewidth, row sep=-\pgflinewidth,
			nodes={draw,
				minimum height=#1,
				anchor=center,
				text width=#1,
				text=white,
				align=center,
				inner sep=0pt
			},
		},
		square matrix/.default=0.45cm
	}
	
	\matrix[square matrix]
	{
		|[fill=black]| $\tau$
		&|[fill=black]|\normalsize 1 &|[fill=black]|\normalsize 2 &|[fill=black]|\normalsize 3 &|[fill=black]|\normalsize 4 &|[fill=black]|\normalsize 5 &|[fill=black]|\normalsize 6 &|[fill=black]|\normalsize 7 &|[fill=black]|\normalsize 8 &|[fill=black]|\normalsize 9 &|[fill=black]|\normalsize 10 &|[fill=black]|\normalsize 11 &|[fill=black]|\normalsize 12 &|[fill=black]|\normalsize 13 &|[fill=black]|\normalsize 14 &|[fill=black]|\normalsize 15 &|[fill=black]|\normalsize 16 &|[fill=black]|\normalsize 17 &|[fill=black]|\normalsize 18 &|[fill=black]|\normalsize 19 &|[fill=black]|\normalsize 20 &|[fill=black]|\normalsize 21 &|[fill=black]|\normalsize 22 &|[fill=black]|\normalsize 23 &|[fill=black]|\normalsize 24 &|[fill=black]|\normalsize 25 &|[fill=black]|\normalsize 26 &|[fill=black]|\normalsize 27 &|[fill=black]|\normalsize 28 &|[fill=black]|\normalsize 29 &|[fill=black]|\normalsize 30 &|[fill=black]|\normalsize 31 &|[fill=black]|\normalsize 32 &|[fill=black]|\normalsize 33 &|[fill=black]|\normalsize 34 &|[fill=black]|\normalsize 35 &|[fill=black]|\normalsize 36 &|[fill=black]|\normalsize 37\\
		|[fill=black]|\normalsize1 & |[fill=red]| &~&~&~&~&~&~&~&~&~&~&~&|[fill=red]|&~&~&~&|[fill=red]|&~&~&~&|[fill=red]|&~&~&~&~&~&~&~&~&~&~&~&~&~&~&~&|[fill=red]|&\\
		|[fill=black]|\normalsize2&~&|[fill=red]|&~&~&~&~&~&~&~&~&~&~&|[fill=red]|&~&~&~&~&~&~&~&~&|[fill=red]|&|[fill=red]|&~&~&~&~&~&~&~&~&~&~&~&~&~&~&\\
		|[fill=black]|\normalsize3&~&~&|[fill=red]|&~&~&~&~&~&~&~&~&~&|[fill=red]|&|[fill=red]|&~&~&~&~&~&~&~&~&~&~&~&~&~&|[fill=red]|&~&~&~&~&|[fill=red]|&~&|[fill=red]|&~&~&\\
		|[fill=black]|\normalsize4&~&~&~&|[fill=red]|&~&~&~&~&~&~&~&~&~&~&~&~&~&~&|[fill=red]|&~&~&~&~&~&~&~&~&|[fill=red]|&~&~&~&~&~&~&~&~&~&\\
		|[fill=black]|\normalsize5&~&~&~&~&|[fill=red]|&~&~&~&~&~&~&~&~&~&~&|[fill=red]|&~&~&~&~&~&~&|[fill=red]|&~&~&~&~&~&|[fill=red]|&~&~&~&~&~&~&~&~&\\
		|[fill=black]|\normalsize6&~&~&~&~&~&|[fill=red]|&~&~&~&~&~&~&~&~&~&~&~&~&~&~&~&~&~&~&~&~&~&~&~&~&~&~&~&~&~&~&~&\\
		|[fill=black]|\normalsize7&~&~&~&~&~&~&|[fill=red]|&~&~&~&~&~&~&~&~&~&~&~&~&~&~&~&~&~&~&~&~&~&~&~&~&~&~&~&~&~&~&\\
		|[fill=black]|\normalsize8&~&~&~&~&~&~&~&|[fill=red]|&~&~&~&~&~&~&~&~&~&~&~&~&~&~&~&|[fill=red]|&~&~&~&~&~&~&~&~&~&~&~&~&~&\\
		|[fill=black]|\normalsize9&~&~&~&~&~&~&~&~&|[fill=red]|&~&~&~&~&~&~&~&~&~&~&~&~&~&~&~&~&~&|[fill=red]|&~&~&~&|[fill=red]|&|[fill=red]|&~&|[fill=red]|&~&~&~&\\
		|[fill=black]|\normalsize10&~&~&~&~&~&~&~&~&~&|[fill=red]|&~&~&~&~&~&~&~&|[fill=red]|&~&~&~&~&~&~&|[fill=red]|&~&|[fill=red]|&~&~&~&~&~&~&|[fill=red]|&~&~&~&\\
		|[fill=black]|\normalsize11&~&~&~&~&~&~&~&~&~&~&|[fill=red]|&~&~&~&~&~&~&~&~&~&~&~&~&~&~&~&~&~&~&~&~&~&~&~&~&~&~&\\
		|[fill=black]|\normalsize12&~&~&~&~&~&~&~&~&~&~&~&|[fill=red]|&~&~&~&~&~&~&~&~&~&~&~&~&~&|[fill=red]|&|[fill=red]|&|[fill=red]|&~&~&~&~&~&~&~&~&~&\\
	};
	
	\end{tikzpicture}
	\caption{Technique Combinations for Honeypots}\label{fig:combinations}
\end{figure*}
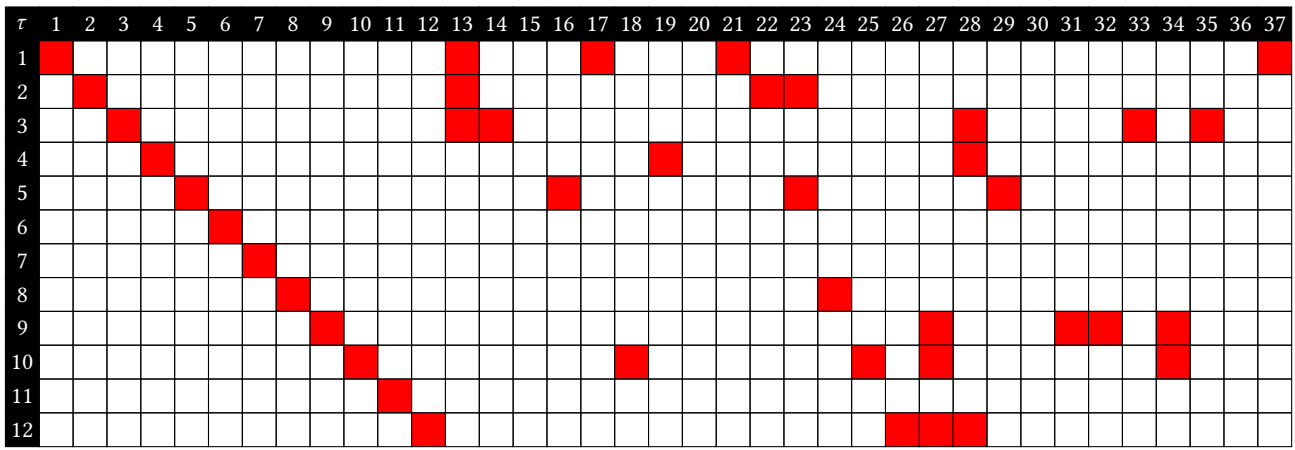

Across the two types of techniques, we identified a total of 37 different techniques. Some early evaluation identified that users who exclusively produce content, without ever following, retweeting or \emph{favouriting} other content, produce little social interaction on Twitter. For this reason, this study only tested the content generating techniques in combination with the content consumption techniques. In addition to combinations of techniques, we also tested the twelve content consumption techniques in isolation to determine the overall effectiveness of these techniques in isolation of the content generation. Ideally, more combinations of these techniques would be tested, but due to restrictions and limitations of Twitter, we were only able to consider 41 different combinations. These combinations are shown in Figure \ref{fig:combinations}. Additionally, some techniques were not tested at all due to the fact that Twitter identified some honeypots as suspicious and suspended some of their user accounts. Additionally, some combinations were tested more than once, at random, to test the repeatability of our findings.

Each of the techniques were constructed as an interactive automated agent using Python and was tested for a period of one week. Execution of the agents on Twitter was largely without incident, except that a number of the original combinations were suspended or terminated by Twitter for suspicious behaviour. We only report on those techniques which survived all the way though our experimentation.

\subsection{Evaluation}
In order to determine the overall effectiveness of the considered techniques, we evaluated the followers that each honeypot attracted. We will look at both the quantitative and qualitative aspects of the attracted accounts. We will record for each honeypot and technique the number of interactions. Each interaction will then be analysed in two ways to gain some qualitative insight. Firstly, we will classify each attracted user as automated agent or human. This will enable us to measure the precision of each technique and honeypot. By separating the techniques we would also be able to get a crude estimate of the recall ability of each technique. Secondly, we will look at the profile characteristics of all the attracted bots.

\section{Results}\label{sec:results}
This section offers the results of the honeypots. We first, review the number of interactions we received per technique and honeypot. We then offer a quantitative review of the users that were attracted by the honeypots.

\subsection{Interactions}
This section reports on the number of interactions each technique and honeypot attracted.
We also include the classification of each attracted user as \textit{human} or \textit{automated} in order to compare the effectiveness of the honeypots at attracting automated agents.

To classify the users, we used a publicly available tool to classify each profile which interacted with the honeypots. We checked all the unique identifiers of the attracted profiles through \texttt{Botometer} which gives a score for each identifier between 0 and 1, where 1 indicates that the user is an automated agent, and 0 a human.\footnote{We used the API interface. The public user interface is available at \url{https://botometer.iuni.iu.edu}}

It is necessary to consider that Botometer itself may not return a perfect result, but was used as a generally useful guide in this study. Botometer returns a number of metrics for each user, but we focused on the overall `universal' score, which reflects how likely a user is to be an automated agent. For the purposes of evaluating our honeypots, we simply considered all users with a percentage score of 50\% or more to be an automated agent, while being aware that not all of these will be.

\begin{figure}
	\includegraphics[width=\linewidth]{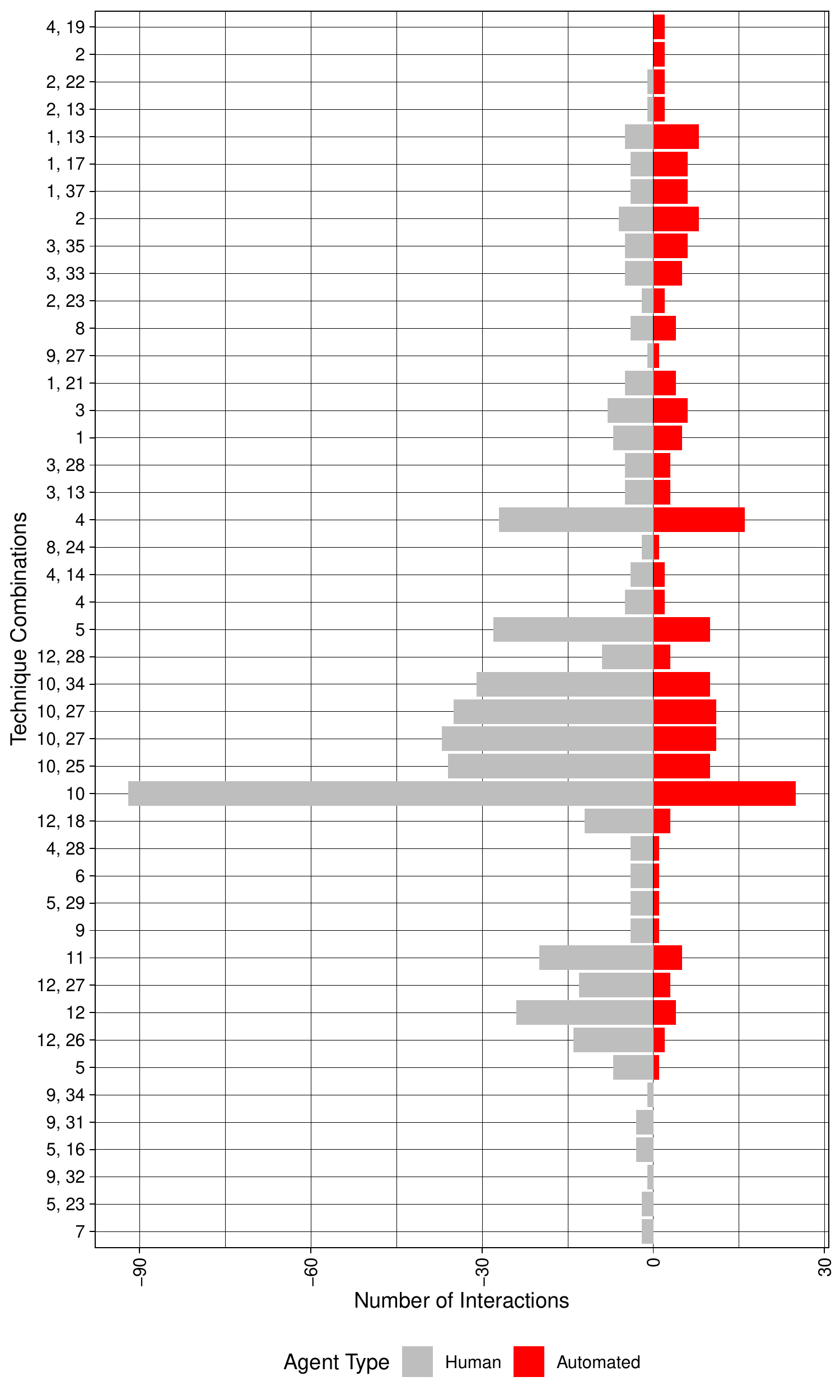}
	\caption{Agent Interactions per Honeypot}
	\label{F:IPHP}
\end{figure}

% latex table generated in R 3.5.0 by xtable 1.8-2 package
% Sun Jul  1 18:29:47 2018
\begin{table*}[]
	\centering
	\caption{Botometer Summary Statistics of the Interacted Users per Technique}
	\label{T:BotSumStats}
	\begin{tabular}{rlrrrrrrrrrr}
		\toprule
		$\tau$ & N & Mean & Median & St. Dev. & Min. & 1\textsuperscript{st} Quartile & 3\textsuperscript{rd} Quartile & Max. & N (Bots) & N (Humans) \\
		\midrule
		1 &  54 & 0.59 & 0.54 & 0.26 & 0.22 & 0.41 & 0.89 & 0.97 &  29 &  25 \\
		2 &  26 & 0.63 & 0.76 & 0.31 & 0.10 & 0.43 & 0.95 & 0.98 &  16 &  10 \\
		3 &  51 & 0.46 & 0.43 & 0.27 & 0.15 & 0.18 & 0.65 & 0.97 &  23 &  28 \\
		4 &  63 & 0.36 & 0.24 & 0.26 & 0.04 & 0.14 & 0.59 & 0.98 &  23 &  40 \\
		5 &  56 & 0.31 & 0.23 & 0.26 & 0.06 & 0.13 & 0.32 & 0.95 &  12 &  44 \\
		6 &   5 & 0.38 & 0.25 & 0.30 & 0.17 & 0.24 & 0.32 & 0.91 &   1 &   4 \\
		7 &   2 & 0.24 & 0.24 & 0.01 & 0.24 & 0.24 & 0.25 & 0.25 &   0 &   2 \\
		8 &  11 & 0.49 & 0.50 & 0.25 & 0.15 & 0.28 & 0.65 & 0.88 &   5 &   6 \\
		9 &  12 & 0.39 & 0.32 & 0.26 & 0.10 & 0.26 & 0.45 & 0.89 &   2 &  10 \\
		10 & 298 & 0.31 & 0.22 & 0.25 & 0.03 & 0.10 & 0.47 & 0.97 &  67 & 231 \\
		11 &  25 & 0.35 & 0.30 & 0.24 & 0.05 & 0.15 & 0.50 & 0.91 &   5 &  20 \\
		12 &  87 & 0.31 & 0.21 & 0.20 & 0.06 & 0.15 & 0.45 & 0.90 &  15 &  72 \\
		13 &  24 & 0.57 & 0.56 & 0.29 & 0.15 & 0.31 & 0.82 & 0.97 &  13 &  11 \\
		14 &   6 & 0.30 & 0.18 & 0.25 & 0.07 & 0.13 & 0.48 & 0.65 &   2 &   4 \\
		16 &   3 & 0.16 & 0.13 & 0.12 & 0.06 & 0.09 & 0.21 & 0.28 &   0 &   3 \\
		17 &  10 & 0.60 & 0.58 & 0.27 & 0.22 & 0.44 & 0.86 & 0.97 &   6 &   4 \\
		18 &  15 & 0.32 & 0.30 & 0.20 & 0.06 & 0.15 & 0.46 & 0.69 &   3 &  12 \\
		19 &   2 & 0.61 & 0.61 & 0.07 & 0.56 & 0.58 & 0.63 & 0.65 &   2 &   0 \\
		21 &   9 & 0.51 & 0.45 & 0.23 & 0.22 & 0.41 & 0.62 & 0.97 &   4 &   5 \\
		22 &   3 & 0.64 & 0.79 & 0.43 & 0.15 & 0.47 & 0.88 & 0.97 &   2 &   1 \\
		23 &   6 & 0.42 & 0.29 & 0.38 & 0.06 & 0.15 & 0.70 & 0.97 &   2 &   4 \\
		24 &   3 & 0.43 & 0.47 & 0.26 & 0.15 & 0.31 & 0.56 & 0.65 &   1 &   2 \\
		25 &  46 & 0.29 & 0.19 & 0.24 & 0.03 & 0.09 & 0.45 & 0.97 &  10 &  36 \\
		26 &  16 & 0.28 & 0.20 & 0.17 & 0.06 & 0.14 & 0.44 & 0.60 &   2 &  14 \\
		27 & 112 & 0.31 & 0.21 & 0.25 & 0.05 & 0.10 & 0.48 & 0.97 &  26 &  86 \\
		28 &  25 & 0.35 & 0.27 & 0.25 & 0.06 & 0.15 & 0.52 & 0.97 &   7 &  18 \\
		29 &   5 & 0.32 & 0.19 & 0.33 & 0.13 & 0.13 & 0.24 & 0.90 &   1 &   4 \\
		31 &   3 & 0.21 & 0.21 & 0.11 & 0.10 & 0.15 & 0.26 & 0.32 &   0 &   3 \\
		32 &   1 & 0.15 & 0.15 &  & 0.15 & 0.15 & 0.15 & 0.15 &   0 &   1 \\
		33 &  10 & 0.48 & 0.52 & 0.28 & 0.15 & 0.20 & 0.65 & 0.97 &   5 &   5 \\
		34 &  42 & 0.30 & 0.21 & 0.24 & 0.05 & 0.10 & 0.48 & 0.97 &  10 &  32 \\
		35 &  11 & 0.50 & 0.58 & 0.29 & 0.15 & 0.22 & 0.70 & 0.97 &   6 &   5 \\
		37 &  10 & 0.60 & 0.58 & 0.27 & 0.22 & 0.44 & 0.86 & 0.97 &   6 &   4 \\ \midrule
		Total & 284 &  &  &  &  &  &  &  &  85 & 199 \\
		\bottomrule
	\end{tabular}
\end{table*}

Through our ensemble of honeypots, we attracted a total of 288 interactions from users on Twitter. These users were then run through the Botometer API. The summary of attracted accounts are shown in Table~\ref{T:BotSumStats}. We see that a total of 284 users could be classified (4 were either deleted, changed to protected, or suspended). Of these 284 accounts, 85 are classified as automated agents and 199 as humans. This means that 30\% the attracted users are actually automated agents, which is well within the scope of previously reported accuracy rates \citep[\textit{cf}][]{Elmendili2017}.

We observe in, Table~\ref{T:BotSumStats}, that, in terms of precision, the best performing technique is $\tau 19$ $(N = 2$, $N_{Bots} = 2$, $N_{Humans} = 0)$, which exclusively attracted automated agents, albeit only two. This technique produces content in gibberish in Xhosa sourced from the content produced by political actors on Twitter.

In terms of recall, we observe that $\tau 10$ $(N = 298$, $N_{Bots} = 67$, $N_{Humans} = 231)$ had by the most interactions but this mostly consists of accounts classified as humans. Recall that $\tau 10$ follows users that post geo-local statuses about a real-time keyword topic. We suspect, but can not show, that the high number of interactions are due to the accounts blindly following back. The method with second highest recall is $\tau 25$ $(N = 2$, $N_{Bots} = 2$, $N_{Humans} = 0)$, which created content in coherent English created from news sources outside of Twitter.

\begin{figure}
	\includegraphics[width=\linewidth]{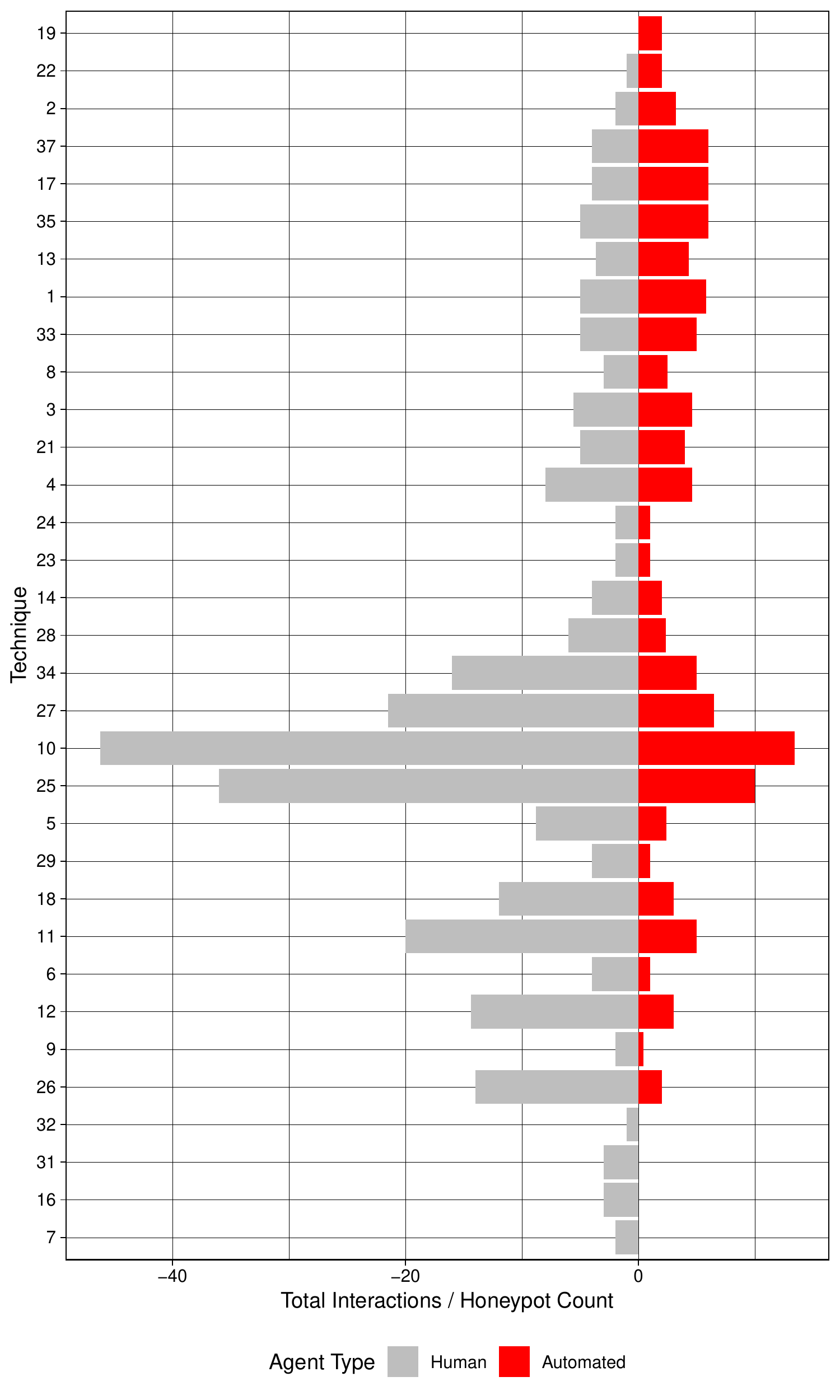}
	\caption{Agent Interactions per Technique}
	\label{F:IPT}
\end{figure}

An overview of interactions we received with each technique is presented in Figure~\ref{F:IPT}, with the interactions for each honeypot (or combination of techniques) illustrated in Figure~\ref{F:IPHP}. The y-axis is ordered from most precise method to least. This means, that the technique or honeypot at the top of the plot managed to attract more automated agents than humans, whereas the technique or honeypot at the bottom attracted more humans than automated agents. The x-axis shows the number of interactions between accounts and honeypots for the given technique. As each technique was used in a number of honeypots, the x-axis is scaled by the overall number of honeypots that made use of the given technique. The number of honeypots used per technique can be deduced from Figure~\ref{fig:combinations}.

In Figure~\ref{F:IPHP} we can see that similar patterns hold, but that with $\tau 10$ the number of human interactions are reduced when it is used in conjunction with other techniques. We also observe that $\tau 1$ is improved by combining it with other techniques. The highest improvement is gained by combining it with $\tau 13$, which increased both recall and accuracy. $\tau 1$ follows a series of political actors, while $\tau 13$ is simply adding content production in English from political sources. The same is true for other languages where combining $\tau 1$ with $\tau 13$ (Afrikaans) and $\tau 37$ (Zulu).

\subsection{Descriptives of Recorded Users}
In order to get an idea of the users that were attracted by our honeypots we review quantitative descriptions of the attracted accounts. Individual descriptions are impractical, so we will provide general descriptives of all attracted accounts.

\begin{figure}
	\includegraphics[width=\linewidth]{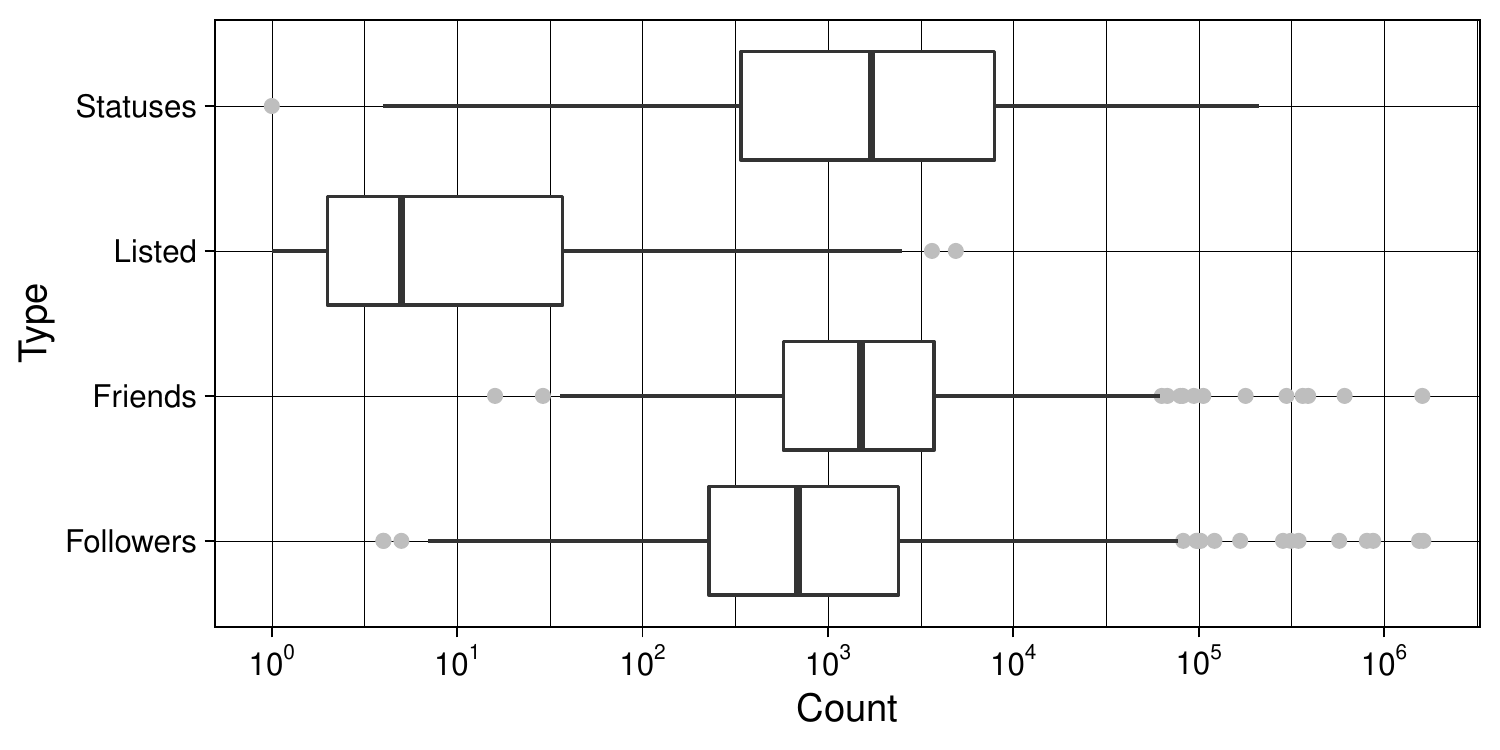}
	\caption{Follower Metrics across all Honeypots}
	\label{F:Box}
\end{figure}

\begin{figure}
	\includegraphics[width=\linewidth]{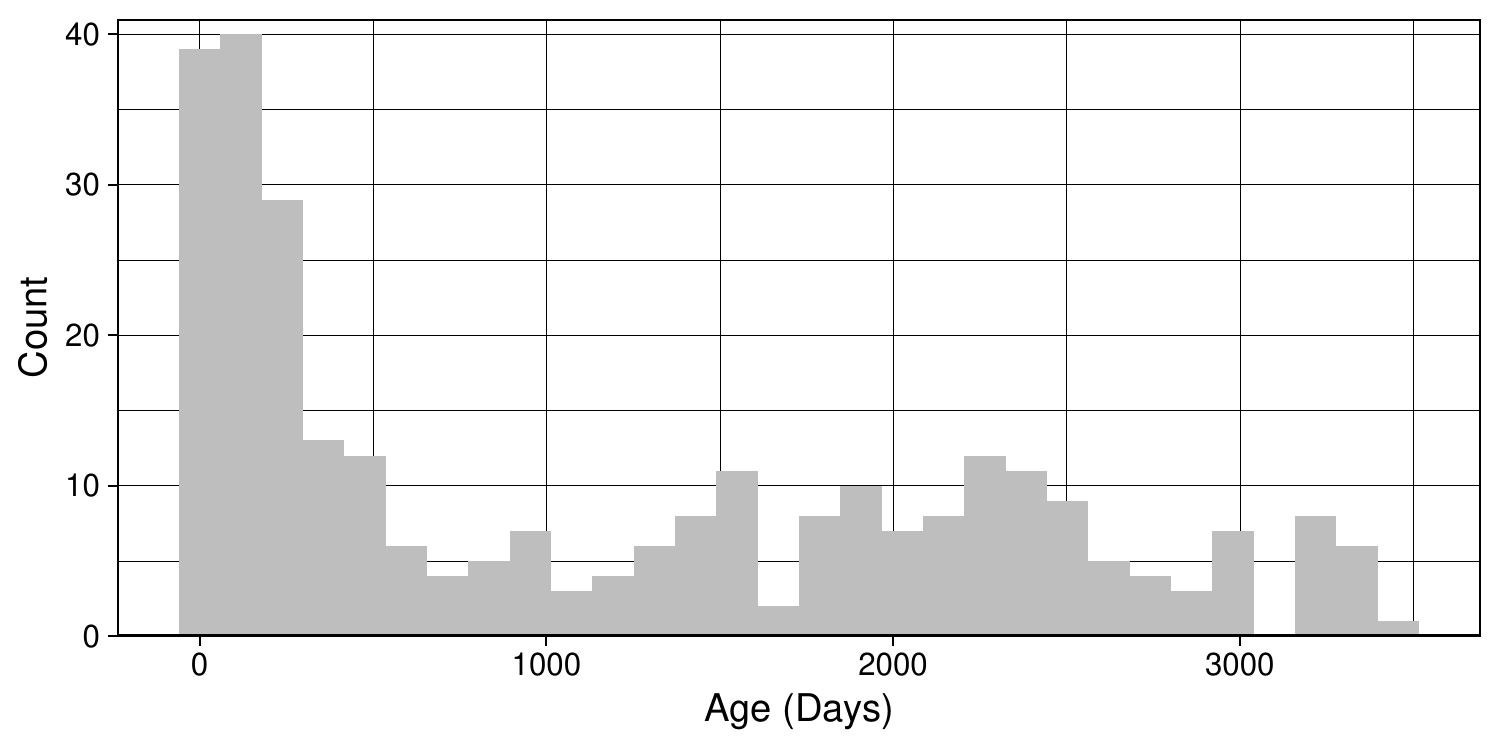}
	\caption{Follower Age across all Honeypots}
	\label{F:Age}
\end{figure}

Table~\ref{T:SumStats} summarises the accounts that interacted with the honeypots. As mentioned, the honeypots attracted a total of 288 accounts. Of these accounts, the average count of statuses, or tweets, is $11,356$, the average number of friends (those an account follows) is $17,312$ and the count of followers (those that follow and account) is $27,927$. What is interesting is that the attracted profiles have on average more followers than friends. However, when considering medians, this is reversed (see Figure~\ref{F:Box}). This is because of a skewed distribution commonly found in social networks. It is also observed that the average age for the attracted accounts is $1,154$ days, with the youngest being $29$ days, and the oldest $3,486$. Figure~\ref{F:Age} indicates that the average age tends to be low. We should also note that we have attracted a total of 10 verified accounts. Verified accounts are accounts that are verified by Twitter as belonging to a public figure or corporation.

\begin{table*}[]
	\centering
	\footnotesize
	\caption{Summary Statistics of Interacted Users}
	\label{T:SumStats}
	\resizebox{0.85\textwidth}{!}{%
		\begin{tabular}{@{\extracolsep{5pt}}lcccccccc}
			\toprule
			Statistic & \multicolumn{1}{c}{N} & \multicolumn{1}{c}{Mean} & \multicolumn{1}{c}{Median} & \multicolumn{1}{c}{St. Dev.} & \multicolumn{1}{c}{Min.} & \multicolumn{1}{c}{1\textsuperscript{st} Quartile} & \multicolumn{1}{c}{3\textsuperscript{rd} Quartile} & \multicolumn{1}{c}{Max.} \\
			\midrule
			Statuses & 288 & 11 356 & 1 666 & 28 433 & 0 & 319 & 7 813 & 210 360 \\
			Friends & 288 & 17 312 & 1496 & 107 871 & 0 & 572 & 3 729 & 1 606 758 \\
			Followers & 288 & 27 927 & 691 & 155 738 & 4 & 228 & 2 400 & 1 628 140 \\
			Listed & 288 & 90 & 1 & 445 & 0 & 0 & 5 & 4 892 \\
			Account Age & 288 & 1 154 & 815 & 1 067 & 29 & 160 & 2 108 & 3 486 \\ \bottomrule
			\multicolumn{8}{l}{\footnotesize{\textbf{Notes:} \textit{Rounded to nearest whole number.}}}\\
		\end{tabular}%
	}
\end{table*}

\section{Conclusion}
The two objectives of this study were to firstly clarify the objectives of social honeypots on OSNs by reiterating the idea that the measure of performance should be recall rather than precision. Unfortunately, there is no way to measure recall in the `wild' since the total number of targets is not known. Nevertheless, the objective is then to maximise the number of accounts attracted by a honeypot. To do this, we isolated the design of honeypots into distinct techniques. Each technique can be tested in isolation and in concert to test for improvements in both recall and precision. We therefore get a sense of recall by comparing the amount of accounts each technique attracted. For instance, we showed that the recall of $\tau 1$ is improved when employed with particular other techniques. Moreover, we can confirm that some techniques far outperform others in attracting interactions, particularly when following other profiles.

This is a distinct improvement over previous studies, where they did not report the honeypot behaviour in isolation, so there is no way to know whether the attraction rate is simply due to `follow back' profiles. Working from these results, it is now possible to systematically build a more contextually effective honeypot, rather than simply designing them with a collection of hypothesised techniques, never knowing which are actually drawing interactions.

We also showed that honeypots which are specifically designed to only attract accounts related to South Africa---by tweeting in local languages---managed to attract interactions. Interestingly, we also managed to attract accounts that are set to Ukrainian and Chinese locales.

For future research, we would design a more systematic phased approach in order to isolate the effectiveness of each technique over time. We would also like to manually inspect the attracted accounts, since we suspect that the standard Botometer measures are conservative and contextually naïve.

This dataset also offers a training dataset for future bot-classification work within the South African context,with the added benefit of describing the context around the attraction of agents, which would help improve the utility of the dataset in machine learning applications.

\balance
\bibliographystyle{ACM-Reference-Format}
\bibliography{CSSG.bib}


\begin{thebibliography}{15}

%%% the Web address.

\ifx \showCODEN    \undefined \def \showCODEN     #1{\unskip}     \fi
\ifx \showDOI      \undefined \def \showDOI       #1{#1}\fi
\ifx \showISBNx    \undefined \def \showISBNx     #1{\unskip}     \fi
\ifx \showISBNxiii \undefined \def \showISBNxiii  #1{\unskip}     \fi
\ifx \showISSN     \undefined \def \showISSN      #1{\unskip}     \fi
\ifx \showLCCN     \undefined \def \showLCCN      #1{\unskip}     \fi
\ifx \shownote     \undefined \def \shownote      #1{#1}          \fi
\ifx \showarticletitle \undefined \def \showarticletitle #1{#1}   \fi
\ifx \showURL      \undefined \def \showURL       {\relax}        \fi
% invisible to TeX
\providecommand\bibfield[2]{#2}
\providecommand\bibinfo[2]{#2}
\providecommand\natexlab[1]{#1}
\providecommand\showeprint[2][]{arXiv:#2}

\bibitem[\protect\citeauthoryear{Bastos and Mercea}{Bastos and Mercea}{2017}]%
        {Bastos2017}
\bibfield{author}{\bibinfo{person}{Marco~T. Bastos} {and} \bibinfo{person}{Dan
  Mercea}.} \bibinfo{year}{2017}\natexlab{}.
\newblock \showarticletitle{{The Brexit Botnet and User-Generated Hyperpartisan
  News}}.
\newblock \bibinfo{journal}{\emph{Social Science Computer Review Preprint}}
  (\bibinfo{year}{2017}), \bibinfo{pages}{1--18}.
\newblock
\showISBNx{0954016157}
\showISSN{15528286}
\urldef\tempurl%
\url{https://doi.org/10.1177/0894439317734157}
\showDOI{\tempurl}


\bibitem[\protect\citeauthoryear{Clark, Williams, Jones, Galbraith, Danforth,
  and Dodds}{Clark et~al\mbox{.}}{2016}]%
        {Clark2016}
\bibfield{author}{\bibinfo{person}{Eric~M. Clark}, \bibinfo{person}{Jake~Ryland
  Williams}, \bibinfo{person}{Chris~A. Jones}, \bibinfo{person}{Richard~A.
  Galbraith}, \bibinfo{person}{Christopher~M. Danforth}, {and}
  \bibinfo{person}{Peter~Sheridan Dodds}.} \bibinfo{year}{2016}\natexlab{}.
\newblock \showarticletitle{{Sifting robotic from organic text: A natural
  language approach for detecting automation on Twitter}}.
\newblock \bibinfo{journal}{\emph{Journal of Computational Science}}
  \bibinfo{volume}{16} (\bibinfo{year}{2016}), \bibinfo{pages}{1--7}.
\newblock
\showISSN{18777503}
\urldef\tempurl%
\url{https://doi.org/10.1016/j.jocs.2015.11.002}
\showDOI{\tempurl}


\bibitem[\protect\citeauthoryear{Elmendili, Maqran, Bouzekri, Idrissi, and
  Chaoui}{Elmendili et~al\mbox{.}}{2017}]%
        {Elmendili2017}
\bibfield{author}{\bibinfo{person}{Fatna Elmendili}, \bibinfo{person}{Nisrine
  Maqran}, \bibinfo{person}{Younes~El Bouzekri}, \bibinfo{person}{El Idrissi},
  {and} \bibinfo{person}{Habiba Chaoui}.} \bibinfo{year}{2017}\natexlab{}.
\newblock \showarticletitle{{A security approach based on honeypots: Protecting
  Online Social network from malicious profiles}}.
\newblock \bibinfo{journal}{\emph{Advances in Science, Technology and
  Engineering Systems Journal}} \bibinfo{volume}{2}, \bibinfo{number}{3}
  (\bibinfo{year}{2017}), \bibinfo{pages}{198--204}.
\newblock
\urldef\tempurl%
\url{https://doi.org/10.25046/aj020326}
\showDOI{\tempurl}


\bibitem[\protect\citeauthoryear{Howard and Kollanyi}{Howard and
  Kollanyi}{2016}]%
        {Howard2016}
\bibfield{author}{\bibinfo{person}{Philip~N. Howard} {and}
  \bibinfo{person}{Bence Kollanyi}.} \bibinfo{year}{2016}\natexlab{}.
\newblock \bibinfo{title}{{Bots, {\#}StrongerIn, and {\#}Brexit: Computational
  Propaganda during the UK-EU Referendum}}.  (\bibinfo{year}{2016}).
\newblock
\showISSN{1556-5068}
\urldef\tempurl%
\url{https://doi.org/10.2139/ssrn.2798311}
\showDOI{\tempurl}


\bibitem[\protect\citeauthoryear{Howard, Woolley, and Calo}{Howard
  et~al\mbox{.}}{2018}]%
        {Howard2018}
\bibfield{author}{\bibinfo{person}{Philip~N. Howard}, \bibinfo{person}{Samuel
  Woolley}, {and} \bibinfo{person}{Ryan Calo}.}
  \bibinfo{year}{2018}\natexlab{}.
\newblock \showarticletitle{{Algorithms, bots, and political communication in
  the US 2016 election: The challenge of automated political communication for
  election law and administration}}.
\newblock \bibinfo{journal}{\emph{Journal of Information Technology and
  Politics}} \bibinfo{volume}{15}, \bibinfo{number}{2} (\bibinfo{year}{2018}),
  \bibinfo{pages}{81--93}.
\newblock
\showISSN{1933169X}
\urldef\tempurl%
\url{https://doi.org/10.1080/19331681.2018.1448735}
\showDOI{\tempurl}


\bibitem[\protect\citeauthoryear{Lee, Caverlee, and Webb}{Lee
  et~al\mbox{.}}{2010a}]%
        {Lee2010a}
\bibfield{author}{\bibinfo{person}{Kyumin Lee}, \bibinfo{person}{James
  Caverlee}, {and} \bibinfo{person}{Steve Webb}.}
  \bibinfo{year}{2010}\natexlab{a}.
\newblock \showarticletitle{{The Social Honeypot Project: Protecting Online
  Communities from Spammers}}. In \bibinfo{booktitle}{\emph{Proceedings of the
  19th international conference on World wide web}}. \bibinfo{address}{Raleigh,
  North Carolina, USA.}, \bibinfo{pages}{1139--1140}.
\newblock
\showISBNx{9781605587998}


\bibitem[\protect\citeauthoryear{Lee, Caverlee, and Webb}{Lee
  et~al\mbox{.}}{2010b}]%
        {Lee2010}
\bibfield{author}{\bibinfo{person}{Kyumin Lee}, \bibinfo{person}{James
  Caverlee}, {and} \bibinfo{person}{Steve Webb}.}
  \bibinfo{year}{2010}\natexlab{b}.
\newblock \showarticletitle{{Uncovering Social Spammers: Social Honeypots plus
  Machine Learning}}. In \bibinfo{booktitle}{\emph{Proceedings of the 33rd
  international ACM SIGIR conference on Research and development in information
  retrieval.}} \bibinfo{address}{Geneva, Switzerland},
  \bibinfo{pages}{435--442}.
\newblock
\showISBNx{9781605588964}
\urldef\tempurl%
\url{https://doi.org/10.1145/1835449.1835522}
\showDOI{\tempurl}


\bibitem[\protect\citeauthoryear{Lee, Eoff, and Caverlee}{Lee
  et~al\mbox{.}}{2011}]%
        {Lee2011}
\bibfield{author}{\bibinfo{person}{Kyumin Lee}, \bibinfo{person}{Brian~David
  Eoff}, {and} \bibinfo{person}{James Caverlee}.}
  \bibinfo{year}{2011}\natexlab{}.
\newblock \showarticletitle{{Seven Months with the Devils: A Long-Term Study of
  Content Polluters on Twitter}}. In \bibinfo{booktitle}{\emph{Icwsm 2011}},
  \bibfield{editor}{\bibinfo{person}{Lada Adamic}, \bibinfo{person}{Ricardo
  Baeza-Yates}, {and} \bibinfo{person}{Scott Counts}} (Eds.).
  \bibinfo{publisher}{The AAAI Press}, \bibinfo{address}{Barcelona},
  \bibinfo{pages}{185--192}.
\newblock
\urldef\tempurl%
\url{https://pdfs.semanticscholar.org/b433/9952a73914dc7eacf3b8e4c78ce9a5aa9502.pdf}
\showURL{%
\tempurl}


\bibitem[\protect\citeauthoryear{Mokube and Adams}{Mokube and Adams}{2007}]%
        {Mokube2007}
\bibfield{author}{\bibinfo{person}{Iyatiti Mokube} {and}
  \bibinfo{person}{Michele Adams}.} \bibinfo{year}{2007}\natexlab{}.
\newblock \showarticletitle{{Honeypots: Concepts, Approaches, and Challenges}}.
  In \bibinfo{booktitle}{\emph{ACM-SE 45 Proceedings of the 45th annual
  southeast regional conference}}. \bibinfo{publisher}{ACM},
  \bibinfo{address}{Winston-Salem, North Carolina, USA},
  \bibinfo{pages}{321--326}.
\newblock
\urldef\tempurl%
\url{https://doi.org/10.1145/1233341.1233399}
\showDOI{\tempurl}


\bibitem[\protect\citeauthoryear{Morstatter, Wu, Nazer, Carley, and
  Liu}{Morstatter et~al\mbox{.}}{2016}]%
        {Morstatter2016}
\bibfield{author}{\bibinfo{person}{Fred Morstatter}, \bibinfo{person}{Liang
  Wu}, \bibinfo{person}{Tahora~H Nazer}, \bibinfo{person}{Kathleen~M Carley},
  {and} \bibinfo{person}{Huan Liu}.} \bibinfo{year}{2016}\natexlab{}.
\newblock \showarticletitle{{A New Approach to Bot Detection: Striking the
  Balance Between Precision and Recall}}. In
  \bibinfo{booktitle}{\emph{International Conference on Advances in Social
  Networks Analysis and Mining}}. \bibinfo{address}{San Francisco, CA, USA},
  \bibinfo{pages}{533--540}.
\newblock
\showISBNx{9781509028467}


\bibitem[\protect\citeauthoryear{Narayanan, Howard, Kollanyi, and
  Elswah}{Narayanan et~al\mbox{.}}{2017}]%
        {Narayanan2017}
\bibfield{author}{\bibinfo{person}{Vidya Narayanan}, \bibinfo{person}{Philip~N
  Howard}, \bibinfo{person}{Bence Kollanyi}, {and} \bibinfo{person}{Mona
  Elswah}.} \bibinfo{year}{2017}\natexlab{}.
\newblock \bibinfo{title}{{Russian Involvement and Junk News during Brexit}}.
  (\bibinfo{year}{2017}), \bibinfo{numpages}{5}~pages.
\newblock


\bibitem[\protect\citeauthoryear{Van~Niekerk}{Van~Niekerk}{2018}]%
        {VanNiekerk2018}
\bibfield{author}{\bibinfo{person}{Brett Van~Niekerk}.}
  \bibinfo{year}{2018}\natexlab{}.
\newblock \showarticletitle{{Information Warfare as a Continuation of Politics:
  An Analysis of Cyber Incidents}}. In \bibinfo{booktitle}{\emph{Conference on
  Information Communications Technology and Society (ICTAS)}}.
\newblock
\showISBNx{9781538610015}


\bibitem[\protect\citeauthoryear{Varol, Ferrara, Davis, Menczer, and
  Flammini}{Varol et~al\mbox{.}}{2017}]%
        {Varol2017}
\bibfield{author}{\bibinfo{person}{Onur Varol}, \bibinfo{person}{Emilio
  Ferrara}, \bibinfo{person}{Clayton~A. Davis}, \bibinfo{person}{Filippo
  Menczer}, {and} \bibinfo{person}{Alessandro Flammini}.}
  \bibinfo{year}{2017}\natexlab{}.
\newblock \showarticletitle{{Online Human-Bot Interactions: Detection,
  Estimation, and Characterization}}. In \bibinfo{booktitle}{\emph{Proceedings
  of the Eleventh International AAAI Conference on Web and Social Media (ICWSM
  2017)}}. \bibinfo{publisher}{Association for the Advancement of Artificial
  Intelligence}, \bibinfo{address}{Montreal, Quebec, Canada},
  \bibinfo{pages}{280--289}.
\newblock
\showISBNx{9781577357889}
\urldef\tempurl%
\url{http://arxiv.org/abs/1703.03107}
\showURL{%
\tempurl}


\bibitem[\protect\citeauthoryear{Wang, Mohanlal, Wilson, Wang, Metzger, Zheng,
  and Zhao}{Wang et~al\mbox{.}}{2012}]%
        {Wang2012}
\bibfield{author}{\bibinfo{person}{Gang Wang}, \bibinfo{person}{Manish
  Mohanlal}, \bibinfo{person}{Christo Wilson}, \bibinfo{person}{Xiao Wang},
  \bibinfo{person}{Miriam Metzger}, \bibinfo{person}{Haitao Zheng}, {and}
  \bibinfo{person}{Ben~Y. Zhao}.} \bibinfo{year}{2012}\natexlab{}.
\newblock \showarticletitle{{Social Turing Tests: Crowdsourcing Sybil
  Detection}}. In \bibinfo{booktitle}{\emph{arXiv preprint}}.
\newblock
\urldef\tempurl%
\url{http://arxiv.org/abs/1205.3856}
\showURL{%
\tempurl}


\bibitem[\protect\citeauthoryear{Yang, Zhang, and Gu}{Yang
  et~al\mbox{.}}{2014}]%
        {Yang2014}
\bibfield{author}{\bibinfo{person}{Chao Yang}, \bibinfo{person}{Jialong Zhang},
  {and} \bibinfo{person}{Guofei Gu}.} \bibinfo{year}{2014}\natexlab{}.
\newblock \showarticletitle{{A taste of tweets: reverse engineering Twitter
  spammers}}. In \bibinfo{booktitle}{\emph{Proceedings of the 30th Annual
  Computer Security Applications Conference}}. \bibinfo{address}{New Orleans,
  USA}, \bibinfo{pages}{86--95}.
\newblock
\showISBNx{9781450330053}
\urldef\tempurl%
\url{https://doi.org/10.1145/2664243.2664258}
\showDOI{\tempurl}


\end{thebibliography}

\end{document}